\documentclass[conference]{IEEEtran}
\IEEEoverridecommandlockouts

\usepackage{graphicx}
\usepackage{csquotes}
\usepackage{scrextend}
\usepackage{hyperref}
\usepackage{comment}
\usepackage{paralist}
\usepackage{microtype}
\usepackage[normalem]{ulem}
\useunder{\uline}{\ul}{}

\usepackage{multirow}
\usepackage[english]{babel}
\usepackage{float}
\usepackage{enumitem}

\usepackage{amsmath}
\usepackage{booktabs}
\usepackage{bm}
\usepackage[table,xcdraw]{xcolor}
\usepackage{microtype}
\usepackage{tikz}
\usepackage{xcolor}
\usepackage{menukeys}

\usepackage[listings,skins,breakable]{tcolorbox}
\newcommand*\circled[1]{\tikz[baseline=(char.base)]{
            \node[shape=circle,fill,inner sep=0.5pt] (char) {\textcolor{white}{#1}};}}

\def\BibTeX{{\rm B\kern-.05em{\sc i\kern-.025em b}\kern-.08em
    T\kern-.1667em\lower.7ex\hbox{E}\kern-.125emX}}
\begin{document}

\title{Automatic ESG Assessment of Companies by Mining and Evaluating Media Coverage Data: NLP~Approach~and~Tool}

\author{
    \IEEEauthorblockN{
        Jannik Fischbach\IEEEauthorrefmark{1}\IEEEauthorrefmark{2}, Max Adam\IEEEauthorrefmark{3}, Victor Dzhagatspanyan\IEEEauthorrefmark{3}, Daniel Mendez\IEEEauthorrefmark{2}\IEEEauthorrefmark{4},  \\ Julian Frattini\IEEEauthorrefmark{4}, Oleksandr Kosenkov\IEEEauthorrefmark{2}, Parisa Elahidoost\IEEEauthorrefmark{2}
    }
    \IEEEauthorblockA{\IEEEauthorrefmark{1} Netlight Consulting GmbH, Munich, Germany, jannik.fischbach@netlight.com}
    \IEEEauthorblockA{\IEEEauthorrefmark{2} fortiss GmbH, Munich, Germany, \{fischbach,mendez,kosenkov,elahidoost\}@fortiss.org}
    \IEEEauthorblockA{\IEEEauthorrefmark{3} Technical University of Munich, Munich, Germany, \{firstname\}.\{lastname\}@tum.de}
    \IEEEauthorblockA{\IEEEauthorrefmark{4} Blekinge Institute of Technology, Karlskrona, Sweden, \{firstname\}.\{lastname\}@bth.se}
}

\maketitle

\begin{abstract}
[Context:] Society increasingly values sustainable corporate behaviour, impacting corporate reputation and customer trust. Hence, companies regularly publish sustainability reports to shed light on their impact on environmental, social, and governance (ESG) factors. [Problem:] Sustainability reports are written by companies and therefore considered a \textit{company-controlled} source. Contrarily, studies reveal that \textit{non-corporate} channels (e.g., media coverage) represent the main driver for ESG transparency. However, analysing media coverage regarding ESG factors is challenging since (1) the amount of published news articles grows daily, (2) media coverage data does not necessarily deal with an ESG-relevant topic, meaning that it must be carefully filtered, and (3) the majority of media coverage data is unstructured. [Research Goal:] We aim to automatically extract ESG-relevant information from textual media reactions to calculate an ESG score for a given company. Our goal is to reduce the cost of ESG data collection and make ESG information available to the general public. [Contribution:] Our contributions are three-fold: First, we publish a corpus of 432,411 news headlines annotated as being environmental-, governance-, social-related, or ESG-irrelevant. Second, we present our tool-supported approach called \texttt{ESG-Miner}, capable of automatically analysing and evaluating corporate ESG performance headlines. Third, we demonstrate the feasibility of our approach in an experiment and apply the \texttt{ESG-Miner} on 3000 manually labelled headlines. Our approach correctly processes 96.7~\% of the headlines and shows great performance in detecting environmental-related headlines and their correct sentiment. 
\end{abstract}


\section{Introduction}
\textbf{Context} Corporate Social Responsibility (CSR) plays an increasingly important role in today's society and affects the public perception of a company. Studies~\cite {Islam21,Schramm16} show that ethically and socially responsible corporate behaviour is positively associated with corporate reputation, customer satisfaction, and customer trust. The necessity of taking social responsibility for corporate actions is also gaining greater recognition among political decision-makers. Specifically, the European Union adopts regulations that force companies to improve their environmental and social impact. Similarly, long-term institutional investors such as insurance companies or pension funds promote CSR~\cite{meng2019institutional}. Driven by the growing interest in CSR, companies strive to integrate social and environmental criteria into their corporate behaviour and release sustainability reports. This trend is particularly evident considering the number of sustainability reports published by S\&P 500 companies. While in 2011 less than 20~\% of these companies reported on their activities towards achieving sustainability goals, this number climbed to 86~\% in 2018~\cite{gillan2021firms}. To make a company's social and environmental impact measurable, the \enquote{Environmental, Social, and Governance} (ESG) concept has evolved. In contrast to evaluating a company solely from a business and financial perspective, ESG analysis considers the impact of a company on the environment and society and evaluates how this impact is managed.

\textbf{Problem} Sustainability reports are written by companies themselves and therefore considered a \textit{company-controlled} source~\cite{margarcia2021}. For this reason, sustainability reports are being criticized since they provide a biased overview of the ESG activities of a company. ESG ratings issued by agencies such as Moody's and Sustainalytics are also subject to critical scrutiny, given that they apply different criteria for ESG assessments, rendering the ratings difficult to compare. Contrarily, non-corporate channels such as media coverage are considered the main driver for ESG transparency~\cite{Hammami2020}. Specifically, companies can hardly control media coverage, making it a more credible and unbiased source of CSR~\cite{Du10,Eisend2011}. However, analysing media coverage regarding ESG criteria is challenging due to the following issues:

\begin{enumerate}
    \item The amount of published news articles grows daily, rendering a manual ESG assessment impractical. This problem becomes even more prominent when aiming for a real-time ESG assessment.
    \item Media coverage data does not necessarily deal with an ESG-relevant topic, meaning it must be carefully filtered.
    \item The majority of media coverage data is \textit{unstructured}, which makes automatic processing increasingly difficult. 
\end{enumerate}

\begin{figure*}[t]
\centering
\fbox{\includegraphics[width=0.9\textwidth]{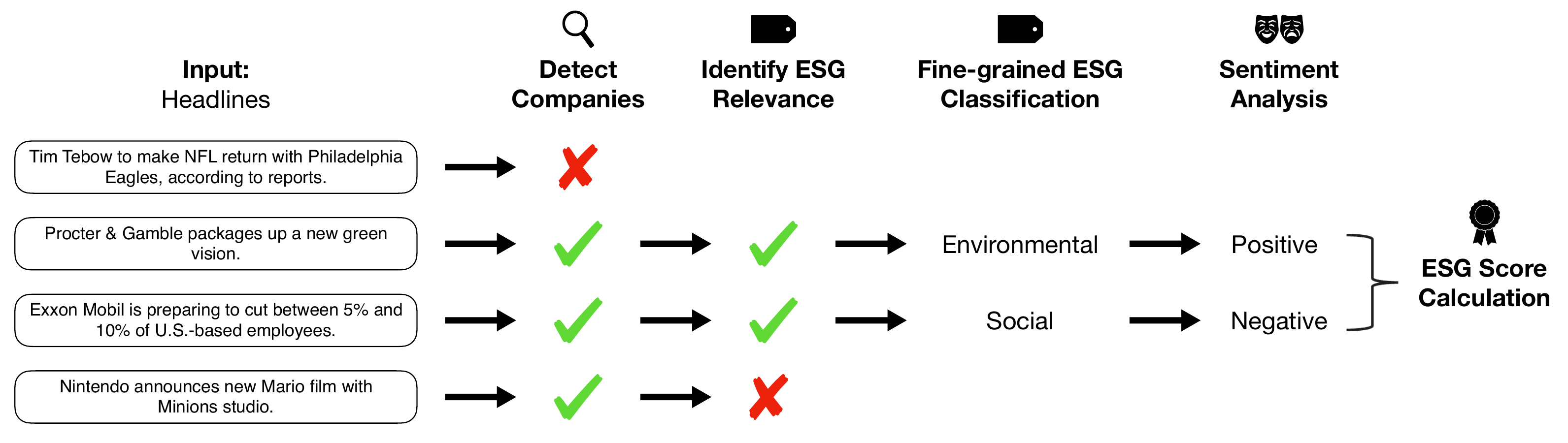}}
\caption{Overview of the \texttt{ESG-Miner} Pipeline.}
\label{fig:esgMinerPipeline}
\end{figure*}

\textbf{Research Goal and Contributions} We aim to extract ESG-relevant information from public, textual media reactions automatically to calculate an ESG score for a given company. We focus on headlines of major news outlets since they represent suitable and credible reactions to corporate-related events. Full-text news articles are an infeasible alternative because of the increased parsing complexity. Since the sentiment of a reputable article is usually conveyed reliably in its title, full-text articles do not promise a significant benefit. As highlighted in \autoref{fig:esgMinerPipeline}, we implement an NLP approach, called \texttt{ESG-Miner}, that is capable of assessing a company's ESG performance in four steps: \circled{1} identify headlines that mention a given company, \circled{2} classify headlines as \enquote{ESG-relevant} or \enquote{ESG-irrelevant}, \circled{3} classify relevant headlines into corresponding ESG category, and \circled{4} identify if the headline is positive/neutral/negative regarding the company's performance in the classified ESG category. Based on the output of this pipeline, we calculate a score representing the public opinion about a company's performance in the three domains. In essence, the \texttt{ESG-Miner} helps to decrease the cost of ESG data collection and makes ESG information available to the general public, enabling consumers and investors to make better-informed purchase- and investment decisions. This paper makes the following contributions (C):

\begin{itemize}
  \item \textbf{C 1}: We publish a data set of 432,411 news headlines annotated as environmental, governance, or social-related. This allows fellow researchers to use the corpus as a benchmark for other ESG-relevant NLP tasks.
  \item \textbf{C 2}: We present our tool-supported approach called \texttt{ESG-Miner}, capable of automatically analyzing and evaluating corporate ESG performance based on headlines. We empirically evaluate our approach in an experiment and compare its prediction performance with human performance. Out of 3000 manually labelled headlines, the \texttt{ESG-Miner} correctly processed 2901 headlines (96.7~\%). In particular, the \texttt{ESG-Miner} performed well in detecting environmental-related headlines along with their correct sentiment.
  \item \textbf{C 3}: To strengthen transparency and facilitate replication, we make our tool, code, and annotated data set publicly available.\footnote[1]{Our replication package can be found at \url{https://doi.org/10.5281/zenodo.7148423}.\label{fnlabel}}
\end{itemize}

\textbf{Related Work} The analysis of news regarding ESG relevance is an active field of research. Yuan~\cite{Yuan21} manually examines Facebook posts by six Chinese and six Korean companies regarding CSR. Guo et al.~\cite{guo20} propose an approach called ESG2Risk that can be used for the volatility prediction of stocks based on ESG news. Luccioni et al.~\cite{luccioni} present ClimateQA, which can analyse sustainability reports to identify climate-relevant sections based on a question-answering approach. Similarly to us, Sokolov et al.~\cite{Sokolov39}, Lee et al.~\cite{Lee22}, and Nugent et al.~\cite{Nugent20} apply language models to detect ESG topics in unstructured text data automatically. However, these approaches only focus on specific intermediate steps (e.g., ESG relevance analysis) and fail to provide an end-to-end pipeline for ESG assessment. Furthermore, the approaches are not published as a tool and thus are inaccessible to the public.

\section{Twitter as a Real-Time Source for News Headlines}
To extract ESG-relevant information from news headlines, we require an annotated set of headlines to train a suitable approach. To the best of our knowledge, no labelled corpus of headlines is available in the research community that could be used for our purposes. Due to the unavailability of adequate data, we created our own training corpus. One way to collect news articles is to download them from news websites using an automated script. However, most news websites are structured differently and use various architectures, requiring us to adjust the script for every website that should be monitored. Contrary, we found that tweets from major news outlets are an easily accessible source of headlines. News outlets usually publish their stories on social media since internet users increasingly access social media to receive news~\cite{social-media-news-source}. 

\textbf{Data Collection and Analysis}
We collected over 552,018 tweets published by @guardian between December 2014 and April 2021. We select \textit{The Guardian} as a data source for news articles due to its general trustworthiness and consistent assignment of tags to articles, on which we base the creation of our data set. However, we plan to expand the data set in further research by adding articles from other news outlets. 78.3~\% of the collected tweets contain a headline and a link to the corresponding article on \url{theguardian.com}. The remaining tweets deal e.g. with promotions and are therefore irrelevant to our use case. Manually mapping the collected headlines to respective ESG categories is not feasible. Thus, we utilize the posted links to collect specific tags assigned to the articles and establish a mapping between tags and ESG categories (see \autoref{tab:tagMapping}). The mapping was generated by the second author and reviewed by the first and third authors. Discrepancies were clarified in extensive discussions to converge at a consistent tag-to-ESG-category mapping. In total, we identified 15,388 different tags.

Each article is associated with at least one tag. When creating the tag-to-ESG category mapping, we focus on the 414 tags assigned to at least 1,000 headlines since analysing all 15,388 tags is too time-consuming. We investigate these 414 tags considering the ESG definition established in literature~\cite{Ziolo19,Sokolov21}. We identified nine tags related to the environmental domain, ten social tags, and five governance-related tags. Articles are often associated with multiple tags (e.g., an environmental-related article is assigned to the tags \enquote{waste}, \enquote{plastics}, and \enquote{recycling}). We treat such articles as only one distinct instance in the respective domain to avoid redundancy in our data set. Therefore, the number of distinct headlines per domain is lower than the number of examples for each domain tag. In rare cases ($\approx$ 3\%), an article was assigned with tags from different ESG categories. We exclude these articles from our data set to prevent issues with training our classifier. 

Following our constructed mapping, the final data set consists of 22,763 distinct articles belonging to the environmental domain, 23,222 social-related, and 4,880 governance-related articles (see \autoref{tab:tagMapping}). In summary, we identified 50,865 distinct articles containing ESG-relevant information. We consider the remaining 381,546 collected articles as ESG-irrelevant. The distribution among the ESG categories is highly imbalanced. While the number of news articles related to the environmental and social domains is balanced, we found significantly fewer governance-related articles. 

\begin{table*}[t]
\caption{Mapping of \textit{The Guardian} Article Tags to ESG Domains. We Present the Tags Assigned to Each Domain, the Number of Guardian News Articles per Tag, and the Number of Distinct Headlines per Domain.}\label{tab:tagMapping}
\centering
\resizebox{0.9\textwidth}{!}{\begin{tabular}{cllllll}
\toprule
\multicolumn{1}{l}{}                                                                         & \multicolumn{6}{c}{\textbf{ESG Categories}}                                                                                                                                                                                               \\ \cmidrule(l){2-7} 
\multicolumn{1}{l}{}                                                                         & \multicolumn{2}{c}{Environmental}                                         & \multicolumn{2}{c}{Social}                                                     & \multicolumn{2}{c}{Governance}                                               \\ \midrule
                                                                                             & \cellcolor[HTML]{E1E2FF}environment      & \cellcolor[HTML]{E1E2FF}21,830 & \cellcolor[HTML]{E1E2FF}gender                 & \cellcolor[HTML]{E1E2FF}3,988 & \cellcolor[HTML]{E1E2FF}corporate governance & \cellcolor[HTML]{E1E2FF}3,798 \\
                                                                                             & climate change                           & 8,209                         & lgbt rights                                    & 2,872                         & tax avoidance                                & 1,528                        \\
                                                                                             & \cellcolor[HTML]{E1E2FF}renewable energy & \cellcolor[HTML]{E1E2FF}1,238  & \cellcolor[HTML]{E1E2FF}discrimination at work & \cellcolor[HTML]{E1E2FF}262   & \cellcolor[HTML]{E1E2FF}executive pay / bonuses        & \cellcolor[HTML]{E1E2FF}751   \\
                                                                                             & waste                                    & 924                          & employment law                                 & 499                         & ethics                                       & 193                           \\
                                                                                             & \cellcolor[HTML]{E1E2FF}plastics         & \cellcolor[HTML]{E1E2FF}675    & \cellcolor[HTML]{E1E2FF}gender pay gap         & \cellcolor[HTML]{E1E2FF}205   & \cellcolor[HTML]{E1E2FF}tax havens           & \cellcolor[HTML]{E1E2FF}592   \\
                                                                                             & recycling                                & 481                            & race                                           & 7,581                         &                                              &                               \\
                                                                                             & \cellcolor[HTML]{E1E2FF}pollution        & \cellcolor[HTML]{E1E2FF}1,908  & \cellcolor[HTML]{E1E2FF}human rights           & \cellcolor[HTML]{E1E2FF}3,081 & \cellcolor[HTML]{E1E2FF}                     & \cellcolor[HTML]{E1E2FF}      \\
                                                                                             & air pollution                            & 676                            & work \& careers                                & 4,106                         &                                              &                               \\
                                                                                             & \cellcolor[HTML]{E1E2FF}ghg emissions    & \cellcolor[HTML]{E1E2FF}2,111  & \cellcolor[HTML]{E1E2FF}job losses             & \cellcolor[HTML]{E1E2FF}2,632 & \cellcolor[HTML]{E1E2FF}                     & \cellcolor[HTML]{E1E2FF}      \\
\multirow{-10}{*}{\textbf{\begin{tabular}[c]{@{}c@{}}tags and\\  \# headlines\end{tabular}}} &                                          &                                & jobs                                & 4                        &                                              &                               \\ \midrule
\multirow{2}{*}{\textbf{\begin{tabular}[c]{@{}c@{}}\# distinct \\ headlines\end{tabular}}}                     & \multicolumn{2}{c}{\textit{22,763}}               & \multicolumn{2}{c}{\textit{23,222}}                    & \multicolumn{2}{c}{\textit{4,880}}    
\\   &  \multicolumn{6}{c}{\textbf{total of ESG-relevant headlines: 50,865}}  \\ 
\bottomrule
\end{tabular}}
\end{table*}

\section{\texttt{ESG-Miner}: Mining and Evaluating Headlines Regarding A Company´s ESG Performance}
This section describes the implementation of the \texttt{ESG-Miner}. As shown in \autoref{fig:esgMinerPipeline}, it is capable of searching for specific company names in headlines (see \autoref{sec:detection}), classifying headlines as ESG-relevant/irrelevant (see \autoref{sec:esgclassification}), demarcating between environmental-, social-, and governance-related headlines (see \autoref{sec:finegrainedesgclassification}), and analyzing the sentiment of headlines (see \autoref{sec:sentimentAnalysis}). The calculation of the final ESG score is described in \autoref{esgScore}. Full tables detailing the results of all experiments performed in this section are included in our replication package.\footref{fnlabel} In all experiments, we utilize a 10-fold \textit{Cross Validation} as studies~\cite{James13} have shown that a model that has been trained this way demonstrates low bias and variance. 

\subsection{Automated Company Detection in News Headlines}\label{sec:detection}
Initially, we need to detect in which headlines a company of interest has been mentioned. It may seem reasonable to use a rule-based system that searches for company names in headlines, e.g. by using regex expressions. However, rule-based systems produce many \textit{False Positives} (FP) and are time-consuming to maintain. We apply a \textit{Named Entity Recognition} approach capable of distinguishing organizations from people, locations, etc. Specifically, we use the spaCy entity recognizer~\cite{EntityRecognizer} to detect organizations in headlines and then use String matching to analyze the names of the identified organizations. For the latter, we use \textit{Term Frequency Inverse Document Frequency} (TF-IDF) to transform company names into a vector and then use a cosine similarity threshold of 0.95 to assess whether the identified organization matches the name of the company of interest. 

\subsection{Automated ESG Classification of News Headlines}
After identifying a company of interest in headlines, we need to determine to which ESG domain the headline belongs to or if the headline is ESG-irrelevant. To this end, we develop an NLP approach consisting of two steps: First, our approach demarcates ESG-relevant from ESG-irrelevant headlines (binary classification). Second, the approach assigns ESG-relevant headlines to one of three ESG classes (multiclass classification). We only use the tags to generate the ground truth data set and deliberately do not base our approach of classifying headlines into ESG categories on the tags assigned by news outlets. Otherwise, the applicability of the \texttt{ESG-Miner} would be limited as it can only process tagged headlines and its performance would strongly depend on the existence and quality of the assigned tags, which might vary among news outlets and might not necessarily be consistent within one outlet. In the following, we describe our applied methods for both steps followed by a report of the results of our experiments.

\subsubsection{Step 1: Demarcating between ESG-relevant and ESG-irrelevant headlines}\label{sec:esgclassification}
The detection of ESG-relevant headlines is a binary classification problem in which we are given a certain headline $\mathcal{X}$ and we are required to produce a nominal label $y \in \mathcal{Y} = \{\text{ESG-relevant}, \text{ESG-irrelevant}\}$. We compare the performance of different approaches for this task: Machine Learning (ML) approaches (e.g., Random Forest) and Deep Learning-based (DL) approaches (e.g., BERT~\cite{devlin19}). 

\textbf{Overview of Applied Approaches} We investigate the use of supervised ML and DL models that learn to predict whether a headline is ESG-relevant based on our labelled data set: Naive Bayes (NB), Support Vector Machines (SVM), Random Forest (RF), Decision Tree (DT), Logistic Regression (LR), and K-Nearest Neighbor (KNN). We use different methods to transform the headlines into sentence embeddings: TF-IDF, doc2vec~\cite{doc2vec}, and spaCy~\cite{spacy2}. We compare the performance of the mentioned ML classifiers with the \textit{Bidirectional Encoder Representations from Transformers} (BERT) model which achieved state-of-the-art results for many NLP tasks (e.g., question answering~\cite{yangQuestionAnswer}, causality detection~\cite{causalityBERT}). BERT is pre-trained on large corpora and can be fine-tuned for any downstream task without requiring much training data. We use the fine-tuning mechanism of BERT and investigate to which extent it can detect ESG-relevant headlines. BERT requires input sequences with a fixed length. Therefore, padding tokens are inserted to adjust all sequences to the same length for headlines that are shorter than this fixed length. Other tokens, such as the classification (CLS) token, are also inserted to provide further information about the headline to the model. For our classification task, we mainly use this token because it stores the information of the whole headline. We feed the pooled information into a single-layer feedforward neural network that uses a softmax layer, which calculates the probability that a sentence is ESG-relevant or not: $\hat y = \text{softmax}(Wv_0 + b)$, where $\hat y$ are the predicted class probabilities for the headline, $W$ is the weighted matrix, $v_0$ represents the first token in the headline (i.e., the CLS token), and $b$ is the bias. We select the class ($c$) with the highest probability as the final classification result: $c = \text{argmax}(\hat y)$.

\textbf{Evaluation Procedure}
Our labelled data set is imbalanced, as only 50,865 headlines are ESG-relevant. We apply \textit{Random Under Sampling} (RUS) to avoid the class imbalance problem. We randomly select headlines from the majority class and exclude them from the data set until a balanced distribution is achieved. From our training corpus, we also exclude headlines that do not refer to companies, as these will already be filtered out in the first step of our pipeline. Further, when training our classifier, we mask company names to prevent bias towards certain companies. Initial experiments have shown that this preprocessing step is essential, given that some companies are significantly more often mentioned in ESG-relevant tweets than others. Our final data set consists of 3,626 headlines, of which 1,813 are equally ESG-relevant and ESG-irrelevant.

\textbf{Experimental Results}
Our experiments reveal that the choice of sentence embeddings significantly influences the performance of all examined ML methods. When using doc2vec, all ML methods show poor performance and deviate significantly from their best-achieved results. For example, NB achieves a macro-F\textsubscript{1} score of 0.77 when using TF-IDF and only a macro-F\textsubscript{1} score of 0.34 when using doc2vec. In contrast, when comparing the performance of the ML methods using TF-IDF and spaCy sentence embeddings, we did not observe any significant performance differences. The only exception is DT, which shows a stronger performance when using TF-IDF compared to spaCy. Interestingly, we found no major performance differences between our ML and DL methods. Although the BERT model possesses a rich linguistic understanding due to its intensive pre-training, it could not outperform most of our ML approaches. In fact, \keys{BERT $\wedge$ Softmax} achieved the same best macro-F\textsubscript{1} score of 0.8 as \keys{TF-IDF $\wedge$ SVM} and \keys{TF-IDF $\wedge$ LR}. \keys{BERT $\wedge$ Softmax} performs significantly better in the detection of ESG-irrelevant headlines than in the detection of ESG-relevant headlines (F\textsubscript{1} score deviation of 27~\%). \keys{TF-IDF $\wedge$ SVM} and \keys{TF-IDF $\wedge$ LR}, on the other hand, show a balanced performance in the detection of both classes. In summary, our experiments confirm the findings of other studies~\cite{FakhouryANKA18,menzies17,causalityBERT,reports21} that applying DL methods does not necessarily lead to better performance compared to conventional ML methods. We found that the ML methods \keys{TF-IDF $\wedge$ SVM} and \keys{TF-IDF $\wedge$ LR} as well as our DL approach \keys{BERT $\wedge$ Softmax} possess the same predictive power according to the macro-F\textsubscript{1} score. Nevertheless, we choose \keys{BERT $\wedge$ Softmax} as our final model for ESG relevance analysis because BERT demonstrated in previous studies~\cite{causalityBERT} that it can handle unseen data well due to its strong language understanding. 

\subsubsection{Step 2: Fine-grained Classification of ESG-relevant Headlines}\label{sec:finegrainedesgclassification}
Fine-grained ESG classification is a multiclass classification problem, in which---given a certain ESG-relevant headline $\mathcal{X}$---we aim to produce a nominal label $y \in \mathcal{Y} = \{\text{environment}, \text{social},\text{governance}\}$. Specifically, the \texttt{ESG-Miner} assigns headlines classified as ESG-relevant to one of the three ESG categories. 

\textbf{Overview of Applied Approaches and Evaluation Procedure}
Similar to the experiments on the binary classification task, we compare the performance of different ML and DL approaches for the fine-grained ESG classification task: NB, SVM, RF, DT, LR, KNN (ML-based approaches) and BERT (DL-based approach). We reuse the 1,813 ESG-relevant headlines included in the balanced data set from step 1 for training our methods for the fine-grained classification of ESG-relevant headlines. We apply RUS to ensure a similar distribution of environmental-, governance-, and social-related headlines. Our final data set consists of 1,794 ESG-relevant headlines, including 598 environmental-, governance-, and social-related headlines each.

\textbf{Experimental Results}
Our previous observation that the choice of sentence embeddings affects the performance of ML methods is corroborated by our experiments for step 2. Using doc2vec leads to the poorest performance of our investigated ML methods. Among the ML methods, \keys{spaCy $\wedge$ RF} achieves the best performance with a macro-F\textsubscript{1} score of 0.86. While our DL approach does not lead to a considerable performance gain in the distinction between ESG-relevant and irrelevant headlines, it outperforms all examined methods in the fine-grained classification of ESG headlines. In fact, \keys{BERT $\wedge$ softmax} can classify ESG-relevant headlines almost flawlessly into the three ESG domains and achieves a macro-F\textsubscript{1} score of 0.95. Hence, we select \keys{BERT $\wedge$ softmax} as the best performing for the second step of your pipeline and integrate it into the \texttt{ESG-Miner}.

\subsection{Sentiment Analysis}\label{sec:sentimentAnalysis}
In the third step, the \texttt{ESG-Miner} analyzes the sentiment of the fine-grained classified headlines. Specifically, it examines whether the company's behaviour mentioned in the headline is considered neutral, positive, or negative regarding the three ESG categories. Several NLP methods are available for analysing the sentiment of NL text~\cite{Birjali21}. However, our initial experiments demonstrated that existing sentiment analysis tools are unsuitable for analysing ESG-relevant tweets. We hypothesize that their poor performance can be attributed to the nature of ESG-relevant headlines that tend to be written objectively. However, conventional sentiment methods rely on subjective statements and keywords to understand the sentiment of a text~\cite{socher13}. Driven by this observation, we implement a new sentiment classifier trained on ESG-relevant headlines.

\textbf{Training Corpus Creation}
We create a training corpus by manually annotating 500 ESG-relevant headlines to train our sentiment classifier. We involve three annotators experienced in the analysis of ESG data and conduct a workshop where we discuss several examples. To ensure consistent annotations, we create an annotation guideline to define the three labels (neutral, positive, and negative) along with a set of sample annotations. We use \textit{Label Studio}~\cite{LabelStudio} for labelling each headline. To verify the reliability of the annotations, we calculate the inter-annotator agreement. We distribute the 500 ESG-relevant headlines among the annotators, ensuring that 100 headlines are labelled by two annotators (overlapping quote of 20~\%). We calculate the Cohen's Kappa~\cite{cohen60} measure based on the overlapping headlines to evaluate how well the annotators can make the same annotation decision for a given category. We interpret Cohen's Kappa using the taxonomy developed by Landis and Koch~\cite{landis77}. Our analysis reveals that in most cases the annotators agreed on the negative labels (kappa: 0.78). We observed moderate agreement on assigning positive (0.58) and neutral labels (0.53). For some samples, the annotators disagreed on whether a headline reports positively on the behaviour of a company or whether the sentiment should still be considered neutral. Across all labels, the mean value is 0.63, which indicates substantial agreement. Therefore, we assess our labelled data set as reliable and suitable for implementing our sentiment classification approach. 

\textbf{Overview of Applied Approaches and Evaluation Procedure}
Sentiment analysis of NL text is a multiclass classifcation problem, in which we are given a certain headline $\mathcal{X}$ and we are required to produce a nominal label $y \in \mathcal{Y} = \{\text{neutral}, \text{positive},\text{negative}\}$. Thus, we face a similar classification problem as for the fine-grained classification of ESG-relevant headlines. We, therefore, use the same methods in our experiments and train them on our annotated data set: NB, SVM, RF, DT, LR, KNN (ML-based approaches) and BERT (DL-based approach). Our annotated data set contains of 125 neutral, 177 negative, and 198 positive headlines. To ensure a similar distribution within the 500 annotated headlines, we apply RUS to avoid the class imbalance problem when training our approaches. Our final data set consists of 375 ESG-relevant headlines, including 125 neutral, positive, and negative headlines each.

\textbf{Experimental Results}
Our results underline the observation from our initial experiments that conventional sentiment methods like CoreNLP~\cite{manning14} and Flair~\cite{akbik2019flair} are not suitable for determining the sentiment of ESG-relevant headlines. Both methods perform poorly and achieve macro-F\textsubscript{1} scores of 0.32 and 0.42. Our self-trained models perform substantially better. For example, a good performance is shown by methods based on TF-IDF sentence embeddings. In fact, \keys{TF-IDF $\wedge$ NB} and \keys{TF-IDF $\wedge$ LR} achieve the best macro-F\textsubscript{1} score of 0.74. Interestingly, \keys{BERT $\wedge$ softmax} cannot cope with this performance and achieves a significantly worse macro-F\textsubscript{1} score of 0.67. 

\subsection{ESG Score Calculation}\label{esgScore}
In the last step, the \texttt{ESG-Miner} summarizes the sentiments of the analyzed headlines and calculates a score for each of the three ESG domains. Each score represents the public opinion about a company's performance in the respective domain. To this end, we apply a simple mean calculation that is also used by sentiment analysis tools~\cite{stine2019sentiment}. We weigh a negative headline by -1, a neutral headline by 0, and a positive headline by +1. Based on this scale, we sum up the weights of all sentiments and divide the result by the total amount of sentiments captured. The resulting score lies between -1 and 1, where a negative value represents a predominantly negative public opinion and vice versa. 

\section{Experiment: \texttt{ESG-Miner} in Practice}
We aim to investigate whether the \texttt{ESG-Miner} is suitable for your use in practice. To this end, we apply it to real-world unseen data and compare its prediction performance with human performance. Specifically, we ask study participants to manually analyze 3000 headlines and compare their analysis results with the one obtained by the \texttt{ESG-Miner}. The participants perform the same steps as the \texttt{ESG-Miner}, i.e. they first search for headlines containing a certain company, then check whether the headlines are ESG-relevant. In our study, we ask the participants to assume the role of a sustainability manager working for ExxonMobil who wants to get an overview of the current public perception of ExxonMobil’s ESG performance. 

\textbf{Study Design}
We randomly collect 3000 headlines posted by \textit{The Independent} (@independent), \textit{Reuters} (@reuters), and \textit{The New York Times} (@nytimes). We involve four study participants and assign 750 headlines to each participant, of which 500 are unique and 250 overlapping. To verify the reliability of their annotations, we calculate the inter-annotator agreement in the same way as in \autoref{sec:sentimentAnalysis}. Checking the inter-annotator agreement is important for two reasons: First, we need to verify whether the participants follow the same understanding during the annotation process and whether the annotations can thus be used as a gold standard to compare human performance with the performance of the \texttt{ESG-Miner}. Second, the inter-annotator agreement gives a first indication on the difficulty of an annotation task and highlights which performance can be expected from an automated approach. In fact, studies~\cite{Berry2021} show that tasks that are challenging for humans also tend to be challenging for machines. 

\begin{figure*}[t]
\centering
\includegraphics[width=0.9\textwidth]{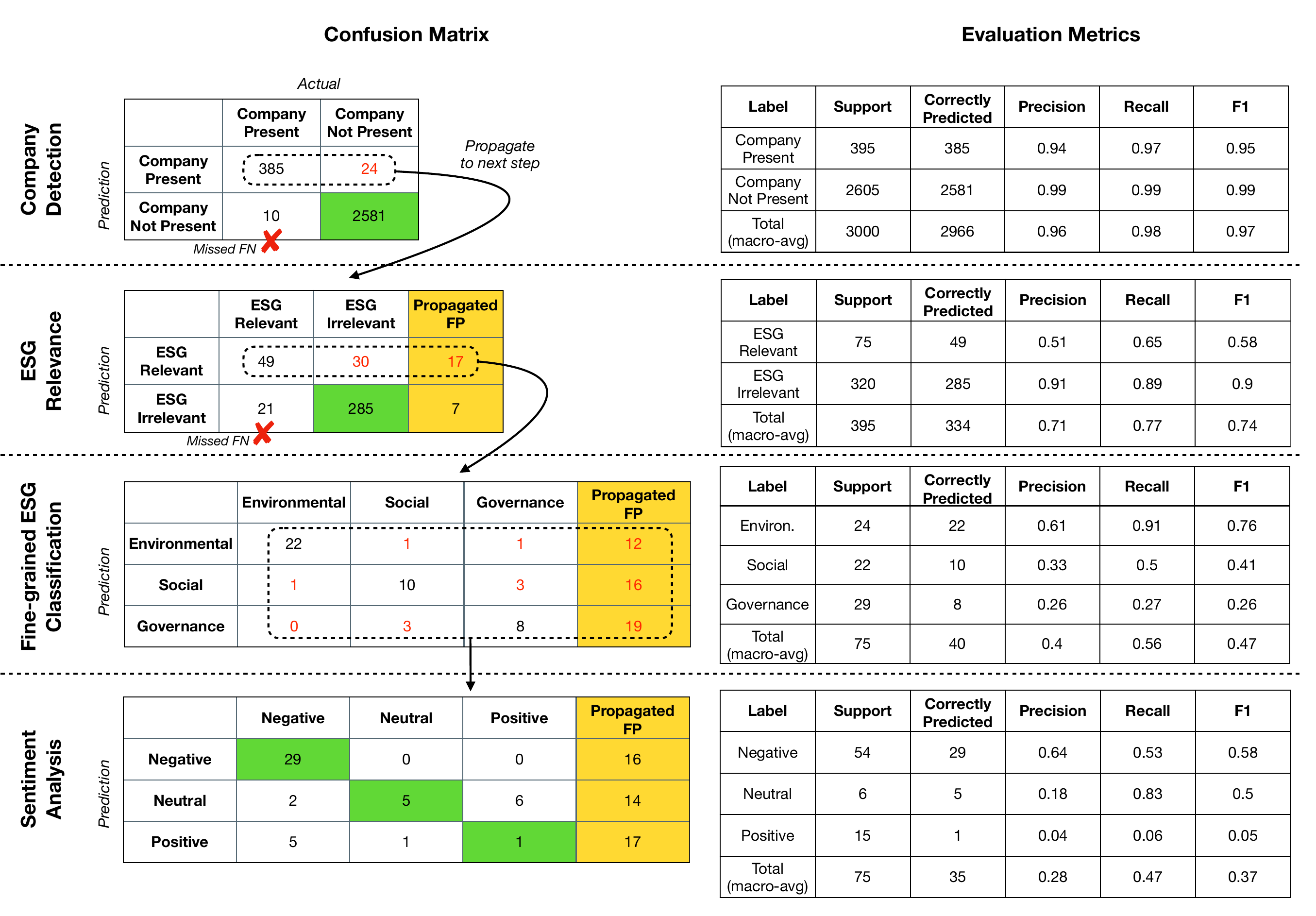}
\caption{Results of Our Case Study: Human Performance vs. \texttt{ESG-Miner}. Propagated FPs Are Highlighted in \textcolor{red}{Red}. Correctly Processed Headlines Are Highlighted in \textcolor{green}{Green}.}
\label{fig:evaluationOverview}
\end{figure*}

\textbf{Study Results}
The study participants performed the identification of companies without any major discrepancies. Out of the 3000 headlines, they identified 395 that refer to ExxonMobil. Our calculated Cohen's kappa value of 0.96 underlines almost perfect agreement among the participants. In case of discrepancies, the study participants discussed the samples to agree on a single label. The \texttt{ESG-Miner} performs very well in recognizing companies and detects 385 of the 395 manually identified headlines. It produces only 24 FP and 10 \textit{False Negatives} (FN), leading to a very strong macro-F$_{1}$ score of 0.97. 

In the second step, the study participants reviewed the 395 headlines concerning ESG relevance. With substantial agreement (kappa: 0.78), they identified 75 ESG-relevant and 320 ESG-irrelevant headlines. In contrast to the manual annotation process where we avoid the propagation of FP and FN through discussions among the participants, the \texttt{ESG-Miner} applies its classifier to all headlines in which it has identified ExxonMobil. Out of these 409 headlines (385 \textit{True Positives} (TP) + 24 FP), it identified 84 headlines as being ESG-relevant. In this context, the \texttt{ESG-Miner} detected 49 of the 75 ESG-relevant headlines, predicted 47 FP, and 21 FN. \autoref{fig:evaluationOverview} demonstrates that the propagation of FPs and the absence of headlines due to being classified as FNs affects the prediction performance causing a significant negative impact on the Precision value. This effect becomes even more prominent in the subsequent steps, since both the FPs resulting from the company detection and newly created FPs are forwarded to the fine-grained ESG classification. Additionally, 21 FNs get lost. 

In the third step, the participants examined the 75 ESG-relevant headlines regarding the three ESG domains. This annotation task raised more misconceptions than the ESG relevance analysis. Across all categories, the annotators achieved only moderate agreement (kappa: 0.59). While the annotators largely agreed on the environmental label, some disagreed on whether a headline is social or governance-related. After resolving these disagreements, the study participants identified 24 environment-related, 22 social-related, and 29 governance-related headlines. The \texttt{ESG-Miner} applied its fine-grained classifier to the 96 headlines which it recognized as being ESG-relevant. Similarly to the human annotator, the \texttt{ESG-Miner} performs well in identifying environmental-related headlines and correctly detects 22 of the 24 headlines manually annotated as environmental-related. However, it shows a significantly worse performance in detecting social-related and governance-related headlines. It recognized only 10 of the 22 social-related headlines and 8 of the 29 governance-related headlines. Nevertheless, \autoref{fig:evaluationOverview} reveals that this performance problem is not attributable to inadequate functioning of our classifier, but is rather caused by the FNs being dropped in preceding steps. Indeed, if we study the performance of the \texttt{ESG-Miner} applied to the 49 ESG-relevant headlines, we obtain a good Accuracy value of 0.81. 

In the last step, the study participants determined the sentiment of the fine-grained classified headlines. Similar to the training corpus creation in \autoref{sec:sentimentAnalysis}, the participants showed substantial agreement in the assignment of negative labels but only moderate agreement regarding positive and neutral labels. 54 of the 75 ESG-relevant headlines were identified as negative, 15 as positive and 6 as neutral. The \texttt{ESG-Miner} correctly detects 29 of the 54 negative headlines and 5 of the 6 neutral headlines. However, it identifies only 1 of the 15 positive headlines. The F\textsubscript{1} scores achieved in this step are also negatively affected by the high number of propagated FPs from the previous steps. The confusion matrix shown in \autoref{fig:evaluationOverview} clearly indicates that the \texttt{ESG-Miner} is certainly capable of determining the sentiment of ESG-relevant headlines with reasonable performance. 

\begin{tcolorbox}[breakable, enhanced jigsaw,arc=0mm,left=1mm,boxsep=1mm,title=\textbf{Summary of Evaluation:}]
\textit{The \texttt{ESG-Miner} performs well in the detection of companies in headlines, the identification of environmental-related headlines as well as in the identification of negatively connoted headlines. Out of all 3000 headlines examined, the \texttt{ESG-Miner} was able to process 2901 headlines (96.7~\%) correctly (see \textcolor{green}{green} highlighting in \autoref{fig:evaluationOverview}): 2581 were correctly filtered out in the company detection step, 285 headlines were correctly identified as ESG-irrelevant, and out of the 75 ESG-relevant headlines, the \texttt{ESG-Miner} classified 35 headlines into the right ESG domain along with the correct sentiment. However, our evaluation also pointed out that the propagation of FPs and FNs leads to performance problems. For example, ESG-irrelevant headlines are erroneously assigned to ESG domains which has to be manually corrected by the user of the \texttt{ESG-Miner}. The results of our case study motivate the further development of the \texttt{ESG-Miner} particularly in terms of improving the ESG relevance analysis to reduce the high number of propagated FPs.}
\end{tcolorbox}

\textbf{Threats to Validity} As in every empirical study, our evaluation is also subject to potential validity threats. This paragraph discusses these threats and describes how we mitigated them. A major threat to internal validity are the annotations by the study participants as an annotation task is to a certain degree subjective. To minimize the bias of the annotators, we assessed the inter-rater agreement. However, imprecise annotations can never be completely avoided. To achieve reasonable generalizability, we selected headlines posted by three different news outlets for the experiment. However, the limited sample size of 3000 headlines does not provide the statistical basis to generalize the results of our experiment beyond the studied samples. Nevertheless, we hypothesize that the \texttt{ESG-Miner} may perform similarly when processing headlines posted by other news outlets as the majority of news outlets use a similar language when formulating headlines. Validation of this claim requires further empirical investigation. 

\section{Conclusion, Limitations, and Outlook}
We take a first step of applying NLP to automatically screening media coverage with respect to ESG categories. We leverage NLP to improve transparency of sustainability-related topics like the ESG performance of specific companies. For this purpose, we create a gold standard corpus of 432,411 headlines annotated as being ESG-irrelevant or environmental-, social-, or governance-related. We use this corpus to train our NLP approach called \texttt{ESG-Miner} capable of detecting companies in headlines, classifying them as ESG-relevant/irrelevant, assigning relevant headlines into one of the three ESG domains, and calculating a final ESG score by analyzing the sentiments of the headlines. Although our evaluation on 3000 unseen headlines showed that the \texttt{ESG-Miner} performs well in detecting environmental-related headlines, we also observed several limitations (L) that need to be addressed in further research:

\begin{itemize}
\item \textbf{L1}: The \texttt{ESG-Miner} is only suitable for analyzing the ESG performance of companies that are frequently featured in news articles. Thus, the ESG performance of companies with little or no coverage in headlines remains intransparent.
\item \textbf{L2}: At present, the \texttt{ESG-Miner} is not able to perform all steps in its pipeline reliably and generates a series of FPs. Hence, it demands human intervention and is only suitable as a semi-automatic method for analyzing a company's ESG performance. In future research, especially the ESG relevance classifier must be improved to prevent the propagation of excessive FPs.
\item \textbf{L3}: The ESG-score calculation currently employs a very rudimentary scoring function. A more sophisticated approach weighting the impact of each relevant headline has to be explored to reflect the companies standing in the respective category more accurately. 
\end{itemize}

\section*{Acknowledgements}
This work was supported by the KKS foundation through the S.E.R.T. Research Profile project at Blekinge Institute of Technology.

\bibliographystyle{IEEEtran}
\bibliography{references}

\begin{thebibliography}{10}
\providecommand{\url}[1]{#1}
\csname url@samestyle\endcsname
\providecommand{\newblock}{\relax}
\providecommand{\bibinfo}[2]{#2}
\providecommand{\BIBentrySTDinterwordspacing}{\spaceskip=0pt\relax}
\providecommand{\BIBentryALTinterwordstretchfactor}{4}
\providecommand{\BIBentryALTinterwordspacing}{\spaceskip=\fontdimen2\font plus
\BIBentryALTinterwordstretchfactor\fontdimen3\font minus
  \fontdimen4\font\relax}
\providecommand{\BIBforeignlanguage}[2]{{%
\expandafter\ifx\csname l@#1\endcsname\relax
\typeout{** WARNING: IEEEtran.bst: No hyphenation pattern has been}%
\typeout{** loaded for the language `#1'. Using the pattern for}%
\typeout{** the default language instead.}%
\else
\language=\csname l@#1\endcsname
\fi
#2}}
\providecommand{\BIBdecl}{\relax}
\BIBdecl

\bibitem{Islam21}
T.~Islam, R.~Islam, A.~H. Pitafi, L.~Xiaobei, M.~Rehmani, M.~Irfan, and M.~S.
  Mubarak, ``The impact of corporate social responsibility on customer loyalty:
  The mediating role of corporate reputation, customer satisfaction, and
  trust,'' \emph{Sustainable Production and Consumption}, vol.~25, pp.
  123--135, 2021.

\bibitem{Schramm16}
H.~Schramm-Klein, J.~Zentes, S.~Steinmann, B.~Swoboda, and D.~Morschett,
  ``Retailer corporate social responsibility is relevant to consumer
  behavior,'' \emph{Business \& Society}, vol.~55, no.~4, pp. 550--575, 2016.

\bibitem{meng2019institutional}
Y.~Meng and X.~Wang, ``Do institutional investors have homogeneous influence on
  corporate social responsibility? evidence from investor investment horizon,''
  \emph{Managerial Finance}, 2019.

\bibitem{gillan2021firms}
S.~L. Gillan, A.~Koch, and L.~T. Starks, ``Firms and social responsibility: A
  review of esg and csr research in corporate finance,'' \emph{Journal of
  Corporate Finance}, p. 101889, 2021.

\bibitem{margarcia2021}
M.~d. Mar Garc{\'i}a-de~los Salmones, A.~Herrero, and P.~Mart{\'i}nez,
  ``Determinants of electronic word-of-mouth on social networking sites about
  negative news on csr,'' \emph{Journal of Business Ethics}, vol. 171, no.~3,
  pp. 583--597, Jul 2021.

\bibitem{Hammami2020}
A.~Hammami and M.~Hendijani~Zadeh, ``Audit quality, media coverage,
  environmental, social, and governance disclosure and firm investment
  efficiency,'' \emph{International Journal of Accounting {\&} Information
  Management}, vol.~28, no.~1, pp. 45--72, Jan 2020.

\bibitem{Du10}
S.~Du, C.~Bhattacharya, and S.~Sen, ``Maximizing business returns to corporate
  social responsibility (csr): The role of csr communication,''
  \emph{International Journal of Management Reviews}, vol.~12, no.~1, pp.
  8--19, 2010.

\bibitem{Eisend2011}
M.~Eisend and F.~K{\"u}ster, \emph{The Effectiveness of Publicity versus
  Advertising: A Meta-Analysis}.\hskip 1em plus 0.5em minus 0.4em\relax
  Wiesbaden: Gabler, 2011, pp. 277--291.

\bibitem{Yuan21}
S.~Yuan, ``Comparing international communication of corporate social
  responsibility by chinese and korean firms on social media,'' \emph{IEEE
  Transactions on Professional Communication}, vol.~64, no.~2, pp. 154--169,
  2021.

\bibitem{guo20}
T.~Guo, N.~Jamet, V.~Betrix, L.-A. Piquet, and E.~Hauptmann, ``Esg2risk: A deep
  learning framework from esg news to stock volatility prediction,'' 2020.

\bibitem{luccioni}
A.~Luccioni, E.~Baylor, and N.~Duchene, ``Analyzing sustainability reports
  using natural language processing,'' 2020.

\bibitem{Sokolov39}
\BIBentryALTinterwordspacing
A.~Sokolov, J.~Mostovoy, J.~Ding, and L.~Seco, ``Building machine learning
  systems for automated esg scoring,'' \emph{The Journal of Impact and ESG
  Investing}, vol.~1, no.~3, pp. 39--50, 2021. [Online]. Available:
  \url{https://jesg.pm-research.com/content/1/3/39}
\BIBentrySTDinterwordspacing

\bibitem{Lee22}
\BIBentryALTinterwordspacing
O.~Lee, H.~Joo, H.~Choi, and M.~Cheon, ``Proposing an integrated approach to
  analyzing esg data via machine learning and deep learning algorithms,''
  \emph{Sustainability}, vol.~14, no.~14, 2022. [Online]. Available:
  \url{https://www.mdpi.com/2071-1050/14/14/8745}
\BIBentrySTDinterwordspacing

\bibitem{Nugent20}
\BIBentryALTinterwordspacing
T.~Nugent, N.~Stelea, and J.~L. Leidner, ``Detecting {ESG} topics using
  domain-specific language models and data augmentation approaches,''
  \emph{CoRR}, vol. abs/2010.08319, 2020. [Online]. Available:
  \url{https://arxiv.org/abs/2010.08319}
\BIBentrySTDinterwordspacing

\bibitem{social-media-news-source}
E.~Shearer, ``Social media outpaces print newspapers in the u.s. as a news
  source,'' \emph{Pew Research Center}, 2018.

\bibitem{Ziolo19}
M.~Ziolo, B.~Z. Filipiak, I.~Bak, and K.~Cheba, ``How to design more
  sustainable financial systems: The roles of environmental, social, and
  governance factors in the decision-making process,'' \emph{Sustainability},
  vol.~11, no.~20, 2019.

\bibitem{Sokolov21}
A.~Sokolov, J.~Mostovoy, J.~Ding, and L.~Seco, ``Building machine learning
  systems for automated esg scoring,'' \emph{The Journal of Impact and ESG
  Investing}, vol.~1, no.~3, pp. 39--50, 2021.

\bibitem{James13}
G.~James, D.~Witten, T.~Hastie, and R.~E. Tibshirani, \emph{An Introduction to
  Statistical Learning}, 2013.

\bibitem{EntityRecognizer}
``spacy entity recognizer,'' \url{https://spacy.io/api/entityrecognizer}.

\bibitem{devlin19}
J.~Devlin, M.-W. Chang, K.~Lee, and K.~Toutanova, ``{BERT}: Pre-training of
  deep bidirectional transformers for language understanding,'' in
  \emph{NAACL'19}.

\bibitem{doc2vec}
\BIBentryALTinterwordspacing
Q.~V. Le and T.~Mikolov, ``Distributed representations of sentences and
  documents,'' \emph{CoRR}, vol. abs/1405.4053, 2014. [Online]. Available:
  \url{http://arxiv.org/abs/1405.4053}
\BIBentrySTDinterwordspacing

\bibitem{spacy2}
\BIBentryALTinterwordspacing
M.~Honnibal and I.~Montani, \emph{spaCy NLP library}, used version: 3.3.
  [Online]. Available: \url{https://spacy.io/}
\BIBentrySTDinterwordspacing

\bibitem{yangQuestionAnswer}
\BIBentryALTinterwordspacing
W.~Yang, Y.~Xie, A.~Lin, X.~Li, L.~Tan, K.~Xiong, M.~Li, and J.~Lin,
  ``End-to-end open-domain question answering with {BERT}serini,'' in
  \emph{Proceedings of the 2019 Conference of the North {A}merican Chapter of
  the Association for Computational Linguistics (Demonstrations)}.\hskip 1em
  plus 0.5em minus 0.4em\relax Minneapolis, Minnesota: Association for
  Computational Linguistics, Jun. 2019, pp. 72--77. [Online]. Available:
  \url{https://aclanthology.org/N19-4013}
\BIBentrySTDinterwordspacing

\bibitem{causalityBERT}
J.~Fischbach, J.~Frattini, A.~Spaans, M.~Kummeth, A.~Vogelsang, D.~Mendez, and
  M.~Unterkalmsteiner, ``Automatic detection of causality in requirement
  artifacts: The cira approach,'' in \emph{Requirements Engineering: Foundation
  for Software Quality}, F.~Dalpiaz and P.~Spoletini, Eds.\hskip 1em plus 0.5em
  minus 0.4em\relax Cham: Springer International Publishing, 2021, pp. 19--36.

\bibitem{FakhouryANKA18}
S.~Fakhoury, V.~Arnaoudova, C.~Noiseux, F.~Khomh, and G.~Antoniol, ``Keep it
  simple: Is deep learning good for linguistic smell detection?'' in
  \emph{International Conference on Software Analysis, Evolution and
  Reengineering, {SANER}}, 2018, pp. 602--611.

\bibitem{menzies17}
W.~Fu and T.~Menzies, ``Easy over hard: A case study on deep learning,'' in
  \emph{Joint Meeting on Foundations of Software Engineering (ESEC/FSE)}, 2017,
  p. 49–60.

\bibitem{reports21}
P.~R. Henao, J.~Fischbach, D.~Spies, J.~Frattini, and A.~Vogelsang, ``Transfer
  learning for mining feature requests and bug reports from tweets and app
  store reviews,'' in \emph{2021 IEEE 29th International Requirements
  Engineering Conference Workshops (REW)}, 2021, pp. 80--86.

\bibitem{Birjali21}
\BIBentryALTinterwordspacing
M.~Birjali, M.~Kasri, and A.~Beni-Hssane, ``A comprehensive survey on sentiment
  analysis: Approaches, challenges and trends,'' \emph{Knowledge-Based
  Systems}, vol. 226, p. 107134, 2021. [Online]. Available:
  \url{https://www.sciencedirect.com/science/article/pii/S095070512100397X}
\BIBentrySTDinterwordspacing

\bibitem{socher13}
\BIBentryALTinterwordspacing
R.~Socher, A.~Perelygin, J.~Wu, J.~Chuang, C.~D. Manning, A.~Ng, and C.~Potts,
  ``Recursive deep models for semantic compositionality over a sentiment
  treebank,'' in \emph{Proceedings of the 2013 Conference on Empirical Methods
  in Natural Language Processing}.\hskip 1em plus 0.5em minus 0.4em\relax
  Seattle, Washington, USA: Association for Computational Linguistics, Oct.
  2013, pp. 1631--1642. [Online]. Available:
  \url{https://aclanthology.org/D13-1170}
\BIBentrySTDinterwordspacing

\bibitem{LabelStudio}
``Label studio,'' \url{https://labelstud.io/}, used version: 1.5.

\bibitem{cohen60}
J.~Cohen, ``A coefficient of agreement for nominal scales,'' \emph{Educational
  and Psychological Measurement}, 1960.

\bibitem{landis77}
J.~R. Landis and G.~G. Koch, ``The measurement of observer agreement for
  categorical data,'' \emph{Biometrics}, 1977.

\bibitem{manning14}
\BIBentryALTinterwordspacing
C.~Manning, M.~Surdeanu, J.~Bauer, J.~Finkel, S.~Bethard, and D.~McClosky,
  ``The {S}tanford {C}ore{NLP} natural language processing toolkit,'' in
  \emph{Proceedings of 52nd Annual Meeting of the Association for Computational
  Linguistics: System Demonstrations}.\hskip 1em plus 0.5em minus 0.4em\relax
  Baltimore, Maryland: Association for Computational Linguistics, Jun. 2014,
  pp. 55--60. [Online]. Available: \url{https://aclanthology.org/P14-5010}
\BIBentrySTDinterwordspacing

\bibitem{akbik2019flair}
A.~Akbik, T.~Bergmann, D.~Blythe, K.~Rasul, S.~Schweter, and R.~Vollgraf,
  ``{FLAIR}: An easy-to-use framework for state-of-the-art {NLP},'' in
  \emph{{NAACL} 2019, 2019 Annual Conference of the North American Chapter of
  the Association for Computational Linguistics (Demonstrations)}, 2019, pp.
  54--59.

\bibitem{stine2019sentiment}
R.~A. Stine, ``Sentiment analysis,'' \emph{Annual review of statistics and its
  application}, vol.~6, pp. 287--308, 2019.

\bibitem{Berry2021}
D.~M. Berry, ``Empirical evaluation of tools for hairy requirements engineering
  tasks,'' \emph{Empirical Software Engineering}, vol.~26, no.~6, p. 111, Aug
  2021.

\end{thebibliography}
\end{document}